\documentstyle
[12pt]
{article}
\begin{document}

\newcommand{\bean}{\begin{eqnarray*}}
\newcommand{\eean}{\end{eqnarray*}}
\newcommand{\ed}{\end{document}}
\newcommand{\pr}{\prime}
\newcommand{\ppr}{\prime\prime}
\newcommand{\cE}{{\tilde E}}
\newcommand{\vphi}{{\varphi}}
\newcommand{\oO}{O(k^{-1})}
\newcommand{\be}{\begin{equation}}
\newcommand{\ee}{\end{equation}}
\newcommand{\barr}{\begin{array}}
\newcommand{\earr}{\end{array}}
\newcommand{\bea}{\begin{eqnarray}}
\newcommand{\eea}{\end{eqnarray}}
\newcommand{\pa}{\partial}
\newcommand{\xx}{\hbox{}^*_*}
\newcommand{\sds}{\subset\hskip - 1em +}

\title{Operator quantization of
constrained WZNW theories and coset constructions.}
\author{A.V.Bratchikov \\
 Kuban State Technological University,\\ 2 Moskovskaya Street,
 Krasnodar, 350072, Russia\\
E-mail:bratchikov@kubstu.ru} \date {December,\,2000}
\maketitle

\begin{abstract}
Using two WZNW theories  for
Lie algebras $g$ and $h,\, h\subset g,$
we construct the associative quotient algebra which includes
a class of $g/h$ coset primary fields and currents.
\end{abstract}

{\bf1.}
In a recent article \cite {B1} for generic $g/h$ coset conformal
field theory \cite {GKO} a class of Virasoro primary
fields and currents was constructed.They are explicitly
expressed in terms of two Wess-Zumino-Novikov-Witten (WZNW) theories
\cite {W,KZ} for Lie algebras $g$ and $h\subset g.$

It turns out that these fields satisfy the usual definition of
primary fields only in the weak sense, i.e. inside correlation
functions.
The object of this paper is to show that outside
correlation functions the primary fields can be represented by elements
of an associative quotient algebra $\Omega/\Upsilon.$
Since coset fields take values in a quotient algebra, the $g/h$
theory is gauge invariant.The gauge transformations are
generated by the ideal $\Upsilon.$

To obtain $\Omega/\Upsilon$ we use broken affine primary
fields of the WZNW theory for $g$ and an auxiliary WZNW theory for $h.$
Elements of the ideal can be treated as first class constraints and our
construction as quantization of the constrained  WZNW theory for
$g\oplus h.$ In the case of Abelian cosets this construction was
obtained in \cite {B3} using a version of the generalized canonical
quantization method \cite {BFF}.

States of coset theories are often identified with
BRST invariant states of gauged WZNW actions \cite {KPSY,HR}.
The algebra $\Omega/\Upsilon $ is also BRST invariant.
The corresponding state
space is equivalent to the ghost free sector of the BRST
approach. However in the full BRST invariant state space which includes
ghosts, the fields under consideration may not satisfy the definition
of primary fields even in the weak sense.

In what follows we treat only the holomorphic part.
\par \smallskip

{\bf 2.}
We begin with the symmetry algebra of the WZNW theory for $g$
\cite {W,KZ}
\bea \label{alg}
[J^a(m),J^b(n)]=if^{abc}J^c(m+n)+{k}m\delta{^{ab}}\delta_{m+n,0},
\eea
\bea \label {vir}
[L^g(m),L^g(n)]=(m-n)L^g(m+n)+c^g\left[\frac 1 {12}
(m^3-m)\right]\delta_{m,-n},
\eea
\bean \label{cr}
[L^g(m),J^a(n)]=-nJ^a(m+n),\qquad
c^g=\frac {2k\,dim\,g} {2k+Q_g}.
\eean
Here $m,n\in Z,$ $f^{abc}$ are the structure constants of $g,$ $k$ and
$c^g$ are the central charges and $Q_g$ is the quadratic
Casimir in the adjoint representation of $g.$
The Virasoro generator
$L^{g}(m)$ is given by Sugawara construction
\bean L^g(m)=\frac {1}
{2k+Q_g}\sum_n{:J^a({m-n})J^a(n):}. \eean

Let $G_R(z)$ be the WZNW primary field
\bean [J^a(m),G_R(z)]=z^{m}G_R(z)t^a_R, \eean
\bean
\label{} [L^g(m),G_R(z)] =z^{m+1}\partial_z G_R(z) +\Delta_R (m+1)z^m
G_R(z),
\eean
\bean
 [t^a_R,t^b_R]=if^{abc}t^c_R,\qquad  \Delta_R=\frac {Q_R} {2k+Q_g}.
\eean
Here $t^a_R$ is the matrix irreducible
representation of the generators of $g$ for the field $G_R(z),$
$\Delta_R$ is the conformal  dimension  of $G_R(z)$ and
${Q_R}$ is the quadratic Casimir of $g$
in the representation $R.$

We shall use the ground state $\vert 0\rangle_g$
of the WZNW theory for $g$
\bea \label {vacg}
\lefteqn {\phantom{1}_g}\,\,\,\,\,\langle0\vert J^a(m\le 0)= J^a(m
\ge 0) \vert 0\rangle_g=0.
\eea

Let $\hat h_k$ be a subalgebra of $\hat g_k.$ One can choose the
basis for $\hat g_k,$ where $\hat h_k$ is generated by
$J^A(m),\,A=1\ldots dim\,h,\,m\in Z.$
Coset Virasoro generators are given by \cite {GKO}
\bea \label {covir}
K(m)=L^g(m)-L^h(m), \qquad m\in Z.
\eea
One can check \cite {GKO} that $K(m)$ commutes with $J^A(n)$
\bea\label {Kaff}
[K(m),J^A(n)]=0
\eea
and satisfy Virasoro algebra (\ref {vir})
with the central charge $c^{g/h}=c^g-c^h.$
The coset primary fields $\phi_i(z)$ are defined by the equations
\bea \label{phi}
[K(m),\phi_i(z)] =z^{m+1}\partial_z
\phi_i(z) +\Delta_i (m+1)z^m \phi_i(z),  \eea
where  $\Delta_i$ is  the conformal dimension of
$\phi_i(z).$

To construct coset fields we extend the operator
algebra of the WZNW theory.
Let us decompose each primary field $G_R$ in
the set of some irreducible representations of $h$ \bea \label {dec}
G_R(z)=\sum_l{G_{R\,l}(z)}=\sum_l{P_lG_{R}(z)}.
\eea
Here $G_{R\,l}(z)$ belongs to the $l'$s representation and $P_l$ is the
corresponding projector.
The field $G_{R\,l}$ satisfies the equation
\bea \label {com1}
[J^A(m),G_{R\,l}(z)]=z^{m}G_{R\,l}(z)t^A_l,
\eea where $t^A_l$ is the representation of the
generators of $h$ for the field $G_{R\,l}(z)$
As well as  $G_{R}(z),$ the field $G_{R\,l}(z)$ is the primary field of
Virasoro algebra (\ref{vir})
\bea \label {1}
\label{L} [L^g(m),G_{R\,l}(z)] =z^{m+1}\partial_z G_{R\,l}(z)
+\Delta_R (m+1)z^m G_{R\,l}(z).
\eea
$G_{R\,l}(z)$ are called broken affine primary fields \cite {HKOC}.
It is convenient to include in
the extended operator algebra all the components
$G_{R\,l}^{\alpha_l}(z)$ of $G_{R\,l}(z).$

To construct coset currents we shall use the field
$J(z)= (J^i(z)), J^i(z)=\sum_mJ^i(m) z^{-m-1},i=dim\,h+1 \ldots  dim\,
g.$ It can be decomposed in the set of some irreducible representations
of $h$ \bean \label {e} J(z)=\sum_s{{J_s}(z)}.
\eean
The field ${J_s}(z)$ satisfies
equation (\ref {com1})
for some $t^A_s$ and
(\ref {L}) with the conformal dimension  $\Delta=1.$ As well as
$G_{R\,l}(z),$  ${J_s}(z)$ is a broken affine primary field.

It follows from (\ref {1}), (\ref {com1}) and similar equations for
${J_s}(z)$  that
\bea \label {KG}
[K(m),G_{R\,l}(z)]&=&
z^{m+1} \left(\partial_z G_{R\,l}(z)- \frac 2  {2k+Q_h}
:J^A(z)G_{R\,l}(z):t^A_l\right) \nonumber \\
\mbox{} &+&\Delta_{R\,l}(m+1)z^m G_{R\,l}(z), \nonumber \\
{[K(m),J_{s}(z)]}&=&
{z^{m+1} \left(\partial_z J_{s}(z)- \frac 2  {2k+Q_h}
:J^A(z)J_{s}(z):t^A_l\right)}          \nonumber \\
\mbox{} &+&{\Delta_{s}(m+1)z^m J_{s}(z)},
\eea
\bea\label{andim}
\Delta_{R\,l}=\Delta_R-\frac {Q_l} {2k+Q_h},\qquad
 \Delta_{s}= 1- \frac {Q_s} {2k+Q_h},
\eea where ${Q_l}\,({Q_s})$ is the quadratic Casimir of $h$ in the
representation $l\,(s).$

At this stage we have the operator algebra
$
{\cal B}_g$ generated by
$J^A(m),G_{R\,l}^{\alpha_l}(z),$
$J_s^{\alpha_s}(z)$ and their modes.
We shall denote by  ${\cal B}_g^{(0)}$ the subalgebra of all the fields
of ${\cal B}_g$ which commute with $J^A(m)$ \bea \label {Bo} [J^A(m),
{\cal B}_g^{(0)}]=0 \eea

The algebra ${\cal B}_g$ is extended further introducing
an auxiliary WZNW theory.
Let $\hat h_{k'}$ be the auxiliary affine Lie algebra
\bean\label{algebra}
[\chi^A(m),\chi^B(n)]=if^{ABC}\chi^C({m+n})+{k'}m
\delta{^{AB}}\delta_{m+n,0},
\eean
where $A,B,C=1\ldots dim\,h,$
$ k'=-k- Q_h$.In our approach the
value of $k'$ is dictated by the conformal invariance of the
$g/h$ theory \cite {B1}.The same $k'$ gives nilpotence of the
corresponding BRST operator \cite {HY}.

Let $\Phi_l$ be the primary
field of the WZNW theory for $\hat h_{k'}$ \bea \label {com2}
[\chi^A(m),\Phi_l(z)]=z^{m}\Phi_l(z)t^{*A}_l, \eea \bea \label {KZ}
\frac {\pa} {\pa z}\Phi_l(z)= \frac {2}{2
k'+Q_h}:\chi^A(z)\Phi_l(z):t^{*A}_l,
 \eea where
$\chi^A(z)=\sum_m{\chi^A(m)z^{-m-1}},\,t^{*A}_l=-(t^{A}_l)^T $ and
$Q_h$ is the quadratic Casimir in the adjoint representation of $h.$
The normal-ordering symbol $:\,:$ means that negative modes of the
currents are on the left and non-negative on the right.
The ground state $\vert 0\rangle_h$ is defined by the relations
\bea    \label {vach}         \lefteqn
{\phantom{1}_h}\,\,\,\,\,\langle0\vert \chi^A(m \le 0)= \chi^A(m \ge 0)
\vert 0\rangle_h=0.
\eea

Let ${\cal B}_{h}$ be the operator algebra
generated by $\chi^A(z),\Phi_l^{\alpha_l}(z)$ and their modes.  Then
the final entended algebra is the direct product ${\cal B}_g\otimes
{\cal B}_{h}.$

\par \smallskip
{\bf 3.}
In
${\cal
B}_g\otimes {\cal B}_{h}$
one can construct the operators
\bea \label{prime}
{\cal G}_{R\,l}(z)= \left (G_{R\,l}(z),\Phi_{l}(z)\right),\quad
{\cal J
}_s(z)= (J_s(z),\Phi_s(z)),
\eea
where $(\cdot,\cdot)$ is the bilinear form
\bea \label {bf4}
\left (G_{R\,l}(z),\Phi_{l}(z)
\right)=
\sum_{\alpha_l}G_{R\,l}^{\alpha_l}(z)\Phi_{l}^{\alpha_l}(z).
\eea

Inside correlation functions ${
\cal G}_{R\,l}(z)$ and ${\cal J
}_s(z)$ are  primary fields of the $g/h$ coset conformal field theory
\cite {B1}.
However they do not satisfy equation
(\ref {phi}).  Instead we have \bea \label {usfe} [K(m),{\cal
G}_{R\,l}(z)]&=& {z^{m+1} \left(\partial_z {\cal G}_{R\,l}(z)+
T_{R\,l}(z)\right) +\Delta_{R\,l}(m+1)z^m {\cal G}_{R\,l}(z)},
\nonumber\\ {[K(m),{\cal J}_s(z)]}&=& {z^{m+1}\left(\partial_z {\cal
J}_s(z)+ T_s(z)\right)+\Delta_s(m+1)z^m {\cal J}_s(z)},
\eea
\bea
T_{R\,l}(z)&=&-\frac {2}{2k+Q_h}\sum_{A=1}^{dim h} {:{\cal
J}^A(z)(G_{R\,l}(z)t^A_{l},\Phi_{l}(z)):},\nonumber \\
T_{s}(z)&=&-\frac {2}{2k+Q_h}\sum_{A=1}^{dim h} {:{\cal
J}^A(z)(J_{s}(z)t^A_{s},\Phi_{s}(z)):}.
\label {T}
\eea
Here  ${\cal J}^A(z)=\sum_m{{\cal J}^A(m)z^{-m-1}},$
${\cal J}^A(m)=J^A(m)+\chi^A(m)$ and :\,: means that negative modes
of ${\cal J}^A(z)$ are on the left and non-negative on the right.
Deriving these equations we used (\ref {KG}),(\ref {KZ}) and the
property of bilinear form (\ref {bf4})
\bean \label {bff}
\left (G_{R\,l}(z)t^A_l,\Phi_{l}(z)
\right)+ \left (G_{R\,l}(z),\Phi_{l}(z)t^{*A}_l\right)=0.
\eean

Using (\ref{Kaff}),(\ref{com1}) and (\ref{com2}) one can check
that
coset Virasoro generators (\ref {covir}) and primary fields (\ref
{prime}) commute with  ${\cal J}^A(m)$
\bean \label {tg} [{\cal J}^A(m),
K(m)]=[{\cal J}^A(m),{\cal G}_{R\,l}(z)]= [{\cal J}^A(m), {\cal
J}_s(z)]=0.  \eean
From this and equation (\ref {usfe}) it follows that
\bea \label {To} [{\cal J}^A(m),T_{R\,l}(z)]=[{\cal J}^A(m),
T_{s}(z)]=0.
\eea
Therefore we can consider equations (\ref {usfe}) in the
subspace $\Omega$ of all the fields of ${\cal B}_g\otimes {\cal B}_{h}$
which commute with ${\cal J}^A(m)$ \bea \label {O} [{\cal
J}^A(m),\Omega]=0.\eea It is easy to see that $\Omega$ is an algebra
with respect to the operator multiplication. In virtue of (\ref {Bo}),
\bea \label {Bo1} {\cal B}_g^{(0)}\subset \Omega.  \eea

The algebra $\Omega$ can be reduced further.
Let us define the subspace
$\Upsilon\subset \Omega $
which is spanned by the fields
\bea                             \label{U}
U(z)=
\sum_{A=1}^{dim\,h}
{:{\cal J}^A(z)U^A(z):},
\eea where $U^A(z)$ are some operators.
It follows from (\ref{T}) and (\ref {To}) that
\bea
\label {Tu}
T_{R\,l},\,T_{s}\in \Upsilon.
\eea

Using equation  (\ref {O}) for $X(z)\in \Omega, U(w)\in \Upsilon,$ we
have \bean \label{id} X(z)U(w)= \sum_{A=1}^{dim\,h}
{:{\cal
J}^A(w)X(z)U^A(w):} .
\eean
Expanding the right-hand side  in powers of
$z-w,$ one can see that all the coefficients are of the form (\ref{U}).
Therefore for $X\in \Omega, U\in \Upsilon$
\bea \label{XU} XU
\in \Upsilon.
\eea
Similarly,
\bea UX \in \Upsilon.
\label{UX}
\eea
From (\ref {XU}) and (\ref {UX}) it follows that $\Upsilon$ is an ideal
of $\Omega$ and the quotient $\Omega/\Upsilon$ is an algebra.Let
$\{X\}\in \Omega/\Upsilon$ be the coset represented by the field $X.$
Then equations (\ref{usfe}) express the fact that $\{{\cal
G}_{R\,l}(z)\},\{{\cal J}_{s}(z)\}$ are  primary fields of the coset
Virasoro algebra \bean [\{K(m)\},\{{\cal G}_{R\,l}(z)\}]&=& z^{m+1}
\partial_z \{{\cal G}_{R\,l}(z)\} +\Delta_{R\,l}(m+1)z^m \{{\cal
G}_{R\,l}(z)\},  \\ {[\{K(m)\},\{{\cal J}_{s}(z)\}]}&=& z^{m+1}
\partial_z \{{\cal J}_{s}(z)\} +\Delta_{s}(m+1)z^m \{{\cal J}_{s}(z)\}.
\eean

Thus we have constructed the operator algebra $\Omega /\Upsilon$ which
includes the coset Virasoro generators $\{K(m)\},$ primary fields
$\{{\cal G}_{R\,l}(z)\} $ and currents $\{{\cal
J}_{s}(z)\}.$

The elements of ${\cal B}_g^{(0)}$
do not depend on auxiliary fields.Therefore, for $X\in {\cal
B}_g^{(0)}$  $\{X\}\ne 0$  and all the fields of ${\cal B}_g^{(0)}$
are in $\Omega/\Upsilon.$

Let us put this in other words.Elements of the ideal
$\Upsilon$ can be treated as constraints.
If $U\in \Upsilon$ we shall write $U\approx 0.$
Since $\Upsilon$ is an algebra, for $U,V\in
\Upsilon $ we have
\bea  \label {subalg} UV \approx 0.
\eea
Therefore  the
constraints are first class (compare with \cite {D}).

Equations (\ref {XU}) and (\ref {UX}) tell us that
all the fields of $\Omega$ are first class \cite {D}.
From this it follows that the theory is invariant with respect to the
left and right gauge transformations \bean \label{} \delta^L_{U(z)}
X(w) =U(z)X(w)\approx 0,\qquad \delta^R_{U(w)}X(z)= X(z)U(w)\approx 0 ,
\eean where $U\in \Upsilon,\,X\in \Omega.$

In virtue  of  (\ref{Tu}),  equations (\ref{usfe}) can be
written in the form (compare with (\ref {phi}))
\bean [K(m),{\cal
G}_{R\,l}(z)]&\approx& z^{m+1} \partial_z {\cal G}_{R\,l}(z)
+\Delta_{R\,l}(m+1)z^m {\cal G}_{R\,l}(z),
\\
\label {ue} {[K(m),{\cal
J}_{s}(z)]}&\approx & z^{m+1} \partial_z
{\cal
J}_{s}(z)
+\Delta_{s}(m+1)z^m
{\cal J}_{s}(z)
\eean

\par \smallskip
{\bf 4.}
The coset ground state $\vert 0\rangle $  can be taken in the form~
$ \vert 0\rangle=\vert 0\rangle_g \otimes\vert 0\rangle_h.$
With respect to the $g/h$ coset Virasoro algebra (\ref {covir})
$\vert 0\rangle$ is  the
$sl(2,C)$ invariant ground state
\bea
K(m)\vert 0\rangle =0, \quad m\ge -1.
\eea
It follows from (\ref {vacg}),(\ref
{vach}) and (\ref {U}) that for $ V(z)\in
\Upsilon$  the correlation function $\langle V(z)\rangle$ vanishes
\bea \label{aver}
\langle V(z)\rangle =0.
\eea

Taking into account (\ref {XU}),(\ref {UX})  and (\ref
{aver}) for $U(z)\in \Upsilon,\,X_i(z)\in \Omega,$ one gets the
gauge Ward identities \bean \langle U(z)X_1(z_1)\ldots X_N(z_N)\rangle
=0.  \eean

From this  and equation (\ref {Tu})
it follows that inside correlation functions
equation (\ref {usfe}) can  be written in the form (\ref {phi}) \bean
[K(m),{\cal G}_{R\,l}(z)]&=& z^{m+1} \partial_z {\cal G}_{R\,l}(z)
+\Delta_{R\,l}(m+1)z^m {\cal G}_{R\,l}(z),
\\
 \label {ue}
{[K(m),{\cal J}_{s}(z)]}&=& z^{m+1} \partial_z {\cal
J}_{s}(z) +\Delta_{s}(m+1)z^m {\cal J}_{s}(z). \eean

In virtue of (\ref {prime}), correlation functions of
coset primary fields and currents can be expressed in terms of two WZNW
theories for Lie algebra $g$ at level $k$ and $h$ at level $k'=-k-Q_h$
\cite {B1}.For instance,
 \bean \langle {\cal G}_{R_1l_1}(z_1)\ldots {\cal
G}_{R_Nl_N}(z_N)\rangle \,\,\,\,\,\,\,\,\,\,\,
\,\,\,\,\,\,\,\,\,\,\,\,\,\,\,\,\,\,\,\,\,\,\,\,\,\,\,\,\,\,\,\,\,\,\,\,\,\,\,\,\,\,\,\,\,\,\,\,\,\,\,\,\,\,\,\,\,\,\,\,\,\,\,\,\,\,\,\,\,\,\
 \eean \bean =\sum_{\alpha_1\ldots
\alpha_N}\lefteqn {\phantom{1}_g}\,\,\,\,\,\langle
G_{R_1l_1}^{\alpha_1}(z_1)\ldots G_{R_Nl_N}^{\alpha_N}(z_N)
\rangle_g\lefteqn {\phantom{1}_h}\,\,\,\,\, \langle
\Phi_{l_1}^{\alpha_1}(z_1) \ldots \Phi_{l_N}^{\alpha_N}(z_N)\rangle_h.
\eean

Let us compare the present approach with the BRST one \cite {KPSY,HR}.
The BRST operator $Q$ can be written in the form
\bea
Q=\int {dz \left (\sum_{A=1}^{dim h}{c_A(z){\cal J}^A(z)}+\tilde
Q(z)\right)} .  \eea Here the ghost field $c_A(z)$ and $\tilde Q(z)$
commute with $\Omega.$ From this and definition (\ref {O}) it follows
that $Q(z)$ commutes with $\Omega$ and our construction is BRST
invariant.  Since we did not use ghost fields the corresponding BRST
invariant state space is ghost free, i.e. has not ghost excitations.

Let us now consider equations (\ref {usfe}) in
the full BRST invariant operator algebra $\Omega'$ which is
constructed using ghost fields.
To divide out the fields $T_{R\,l},\,T_{s}$ one must
construct the ideal of $\Omega'$ which includes these fields and does
not include coset Virasoro generators and primary fields.
$\Upsilon$ is not an ideal of $\Omega'$ and it is not
clear whether such ideal exists.
We leave this problem for future investigations.

In conclusion, we have described the reduction which
leads from the WZNW theory for $g\oplus h$ to an operator (sub)algebra
of the $g/h$ coset conformal field theory. This approach can
be used for operator quantization of more general coset constructions
and other constrained systems.

\bigskip

I am grateful to I.V.Tyutin for a useful discussion.

\end{document}